\documentclass[fleqn,usenatbib]{mnras}

\usepackage{newtxtext,newtxmath}
\usepackage[T1]{fontenc}
\usepackage{ae,aecompl}

\usepackage{graphicx}	% Including figure files
\usepackage{amsmath}	% Advanced maths commands
\usepackage{amssymb}	% Extra maths symbols
\usepackage{bm}% bold math
\usepackage[colorinlistoftodos]{todonotes}
\usepackage{ dsfont }

\DeclareMathOperator{\e}{e}

\DeclareMathOperator{\n}{\text{n}}
\DeclareMathOperator{\p}{\text{p}}
\DeclareMathOperator{\nn}{\text{nn}}
\DeclareMathOperator{\pn}{\text{pn}}
\DeclareMathOperator{\np}{\text{np}}
\DeclareMathOperator{\pp}{\text{pp}}

\title[Force on a vortex]{Force on a neutron quantised vortex pinned to proton fluxoids in the superfluid core of cold neutron stars}

\author[A. Sourie and N.~Chamel]{
Aur\'elien Sourie,$^{1,2}$\thanks{E-mail: asourie@ulb.ac.be}
and Nicolas Chamel$^{1}$\thanks{E-mail: nchamel@ulb.ac.be}
\\
$^{1}$Institut d'Astronomie et d'Astrophysique, CP-226, Universit\'e Libre de Bruxelles, 1050 Brussels,
Belgium\\
$^{2}$Laboratoire Univers et Th\'eories, Observatoire de Paris, PSL Research University, CNRS, Universit\'e Paris Diderot, \\  Sorbonne Paris Cit\'e,  5 place Jules Janssen, 92195 Meudon, France\\
}

\date{Accepted XXX. Received YYY; in original form ZZZ}

\pubyear{2020}

\begin{document}
\label{firstpage}
\pagerange{\pageref{firstpage}--\pageref{lastpage}}
\maketitle

% Abstract of the paper
\begin{abstract}
The superfluid and superconducting core of a cold rotating neutron star is expected to be threaded by a tremendous number of neutron quantised vortices and proton fluxoids. Their interactions are unavoidable and may have important astrophysical implications. In this paper, the various contributions to the force acting on a single vortex to which fluxoids are pinned are clarified. The general expression of the force is derived by applying the variational multifluid formalism developed by Carter and collaborators. Pinning to fluxoids leads to an additional Magnus type force due to proton circulation around the vortex. Pinning in the core of a neutron star may thus have a dramatic impact on the vortex dynamics, and therefore on the magneto-rotational evolution of the star.

\end{abstract}

% Select between one and six entries from the list of approved keywords.
% Don't make up new ones.

\section{Introduction}

Even before their actual discovery, neutron stars (NSs) were expected to be so dense that neutrons and protons in their interior may be in a superfluid state (see, e.g., \cite{chamel2017} and references therein). This theoretical prediction was later confirmed by the very long relaxation times following the first detections of pulsar sudden spin-ups so-called frequency `glitches'~\citep{haskell2015models}. Nucleon superfluidity in the core of NSs has recently found additional support from the direct monitoring of the rapid cooling of the young NS in Cassiopeia A~\citep{page2011rapid, shternin2011cooling}. However, the interpretation of these observations remains controversial~\citep{posselt2018,ho2019}.

Because NSs are rotating, their interior is threaded by a huge number of neutron quantised vortices, each carrying a quantum $\kappa_{\n}= h/(2\, m_{\n}) \simeq 2\times 10^{-3}  \text{ cm}^2~\text{s}^{-1}$ of circulation, where $h$ is the Planck constant and $m_{\n}$ is the neutron rest mass. The mean surface density of vortices is proportional to the angular frequency $\Omega$ and is given by~\citep{chamel2017}
\begin{equation}
\label{surfdens_vortex}
\mathcal{N}_{\n}= 4\,  m_{\n}\,  \Omega / h\simeq 6\times 10^5 / P_{10}\  \text{cm}^{-2}\, ,
\end{equation} 
where $P_{10}= P / 10\text{ ms}$ is the observed rotation period of the neutron star. Assuming that protons in NS cores form a type-II superconductor~\citep{baym1969superfluidity}, the magnetic flux penetrates the stellar interior only via fluxoids, each carrying a quantum magnetic flux $\phi_0=h\, c/(2\, e)\simeq 2\times 10^{-7} \text{ G~cm}^{2}$, where $c$ is the speed of light and $e$ denotes the proton electric charge. For typical NS magnetic fields, the number of fluxoids is considerably larger than that of vortices, their mean surface density being given by~\citep{chamel2017}
\begin{equation}
\label{surfdens_fluxtube}
\mathcal{N}_{\p}  = B/ \phi_0 \simeq 5\times 10^{18}\, B_{12} \ \text{cm}^{-2}\, , 
\end{equation} 
where $B_{12} = B / 10^{12}\text{\ G}$ is the stellar internal magnetic field. Interactions between neutron vortices and proton fluxoids are therefore unavoidable, and are pivotal in the magneto-rotational evolution of NSs. In particular, vortices may pin to fluxoids \citep{muslimov1985vortex,sauls1989superfluidity,srinivasan1990novel,ruderman1998neutron} (see also \cite{alpar2017} for a recent review), and this may have important  implications for various astrophysical phenomena, such as precession~\citep{sedrakian1999precession, link2006incompatibility, glampedakis2008stability}, r-mode instability~\citep{haskell2009rmodes, haskell2014new} and pulsar glitches~\citep{sedrakian1995superfluidII,sidery2009effect, glampedakis2009, haskell2013,haskell2015models, gugercinoglu2017, sourie2017global, haskell2018, graber2018}. However, the detailed force acting on individual vortices to which fluxoids are pinned remains poorly understood. In particular, the contribution associated with the proton circulation induced by pinned fluxoids has been generally overlooked or treated phenomenologically~(see, e.g., \cite{glampedakis2011magnetorot}). 

Building on the recent study of \cite{gusakov2019}, who determined the force acting on a single fluxoid and clarified the role of degenerate electrons, we derive in this paper the general expression for the force per unit length acting on a neutron vortex to which $N_{\p}$ proton fluxoids are pinned. To this end, we follow a general approach originally developed by \cite{carter2002relativistic} in the relativistic framework, and later adapted to the Newtonian context by \cite{carter2005covariant}. The general expression of the vortex velocity is calculated and the role of pinning on the vortex dynamics is discussed. The implications for pulsar glitches are studied in an accompanying paper~\citep{sourie20b}.

\section{Force acting on a single vortex pinned to fluxoids}

\subsection{General definition}

Let us consider a rigid and infinitely long straight neutron superfluid vortex to which $N_{\p}$ proton fluxoids are pinned. The medium in which the vortex is embedded is assumed to be asymptotically uniform, stationary and longitudinally invariant, along say the $z$ axis. 

\begin{figure}
	 \includegraphics[width=3in]{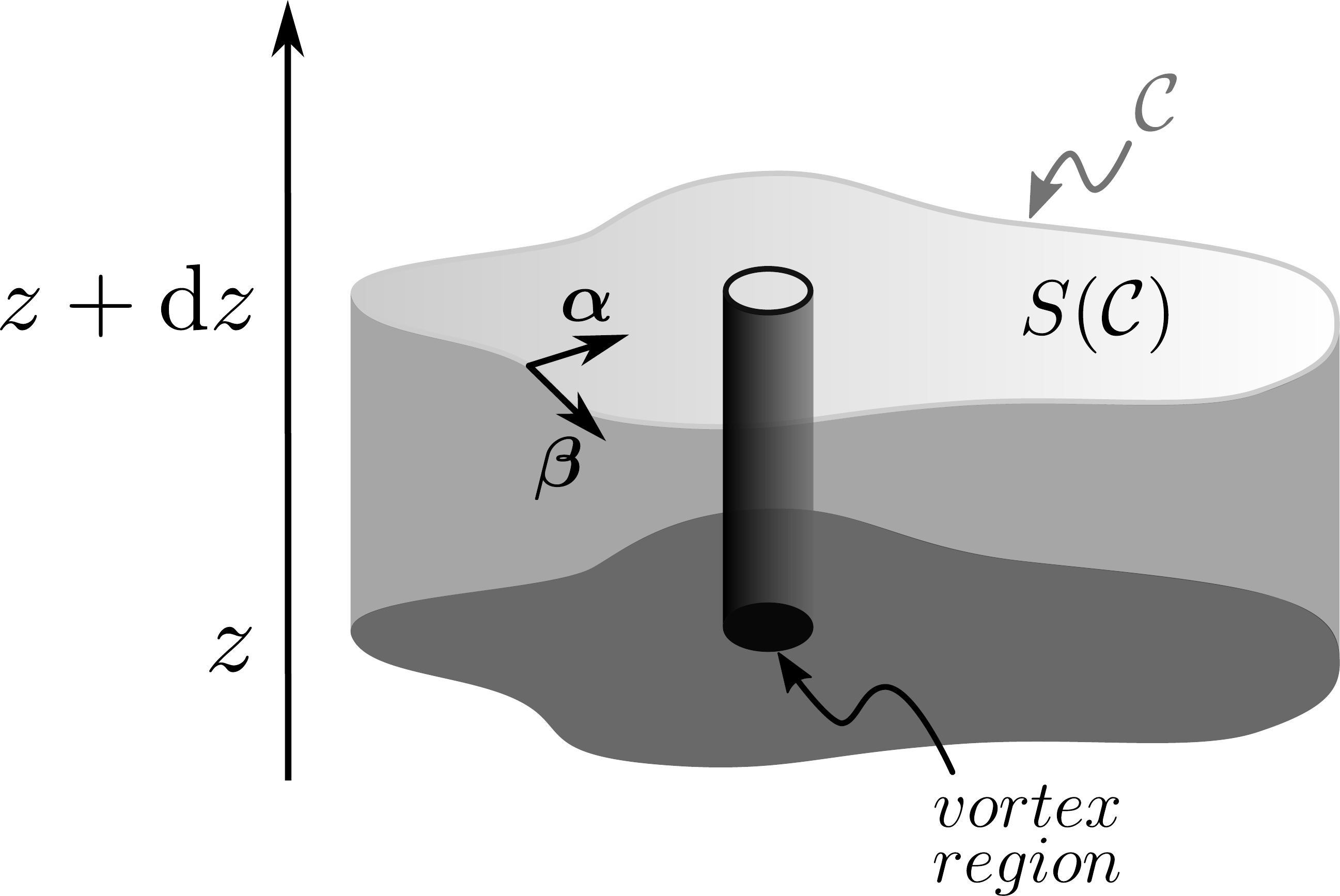}
	\vspace*{8pt}
	\caption{Schematic picture illustrating the fluid element contributing to the force per unit length acting on the vortex directed along the $z$ axis. The parallel sections $S(\mathcal{C})$ lie in the plane perpendicular to this axis. See text for details.}
	\label{fig:force}
\end{figure}

The force density $\pmb{f}$ acting on a matter element is defined by the divergence of the momentum-flux tensor $\Pi^{ij}$ ($i$, $j$ denoting space coordinate indices), 
\begin{equation}
f_i\equiv\nabla_j \Pi^j_i \, .
\end{equation}
The force $\text{d}\pmb{F}$ exerted on a vortex segment of length $\text{d}z$ by a matter element whose volume is delimited by a closed contour $\mathcal{C}$ encircling the vortex and the pinned fluxoids, as represented on Fig.~\ref{fig:force}, is thus given by
\begin{align}\label{def-force}
	\text{d}F_i&=-\iiint f_i\,  \text{d}V=-\iiint \nabla_j \Pi^j_i \, \text{d}V \nonumber \\
	&=\iint_{S(\mathcal{C})} \Pi^j_i(z)\hat{z}_j\,\text{d}S-\iint_{S(\mathcal{C})} \Pi^j_i(z+\text{d}z)\hat{z}_j\, \text{d}S \nonumber \\
	&\ \ +\text{d}z \oint_\mathcal{C} \Pi^j_i\alpha_j \, \text{d}\ell \, ,
\end{align}
where we have made use of Stokes' theorem, and $\pmb{\alpha}$ is a unit vector perpendicular to both the vortex line and the contour $\mathcal{C}$, and is oriented inside the contour. Longitudinal invariance along the vortex line implies that $\Pi^{ij}$ is independent of $z$. The two surface integrals in the second line of Eq.~\eqref{def-force} thus cancel each other. The force per unit length acting on the vortex and the pinned fluxoids can be finally expressed as 
\begin{equation}
\label{jouko-def}
\mathcal{F}_{i}\equiv\frac{\text{d}F_i}{\text{d}z}=\oint_\mathcal{C} \Pi^j_i\, \alpha_j \, \text{d}\ell \, .
\end{equation}

\begin{figure}
    \centering
	 \includegraphics[width=2.5in]{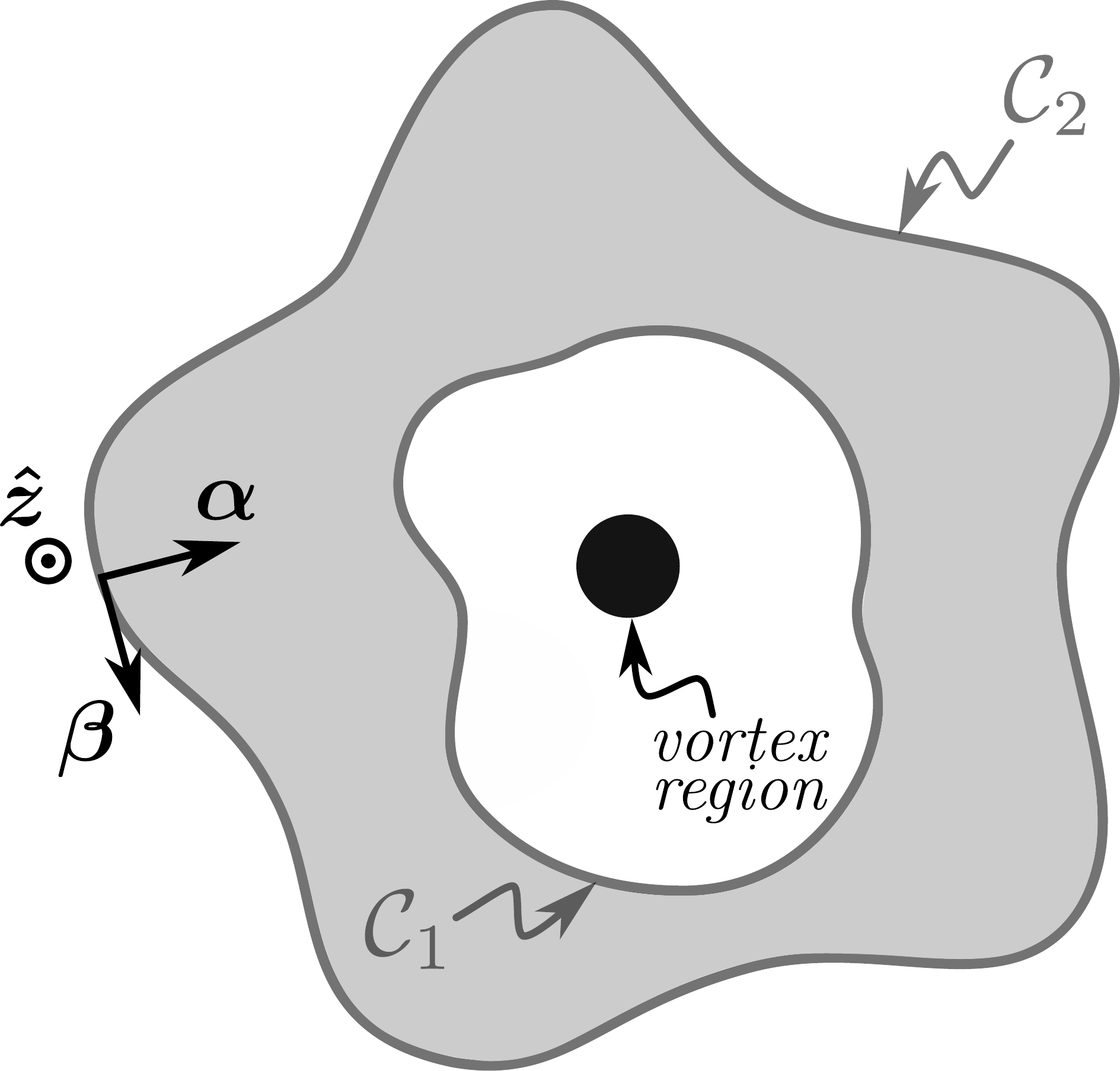}
	\vspace*{8pt}
	\caption{Schematic picture illustrating the surface $\mathcal{S}(\mathcal{C}_2)\, \backslash\, \mathcal{S}(\mathcal{C}_1)$ (shaded area) delimited by two different contours $\mathcal{C}_1$ and $\mathcal{C}_2$ around the vortex region.}
	\label{fig:surface}
\end{figure}

The force \eqref{jouko-def} is well-defined provided the contour integral is evaluated at sufficiently large distances from the vortex where the force density vanishes, $f_i=0$. Indeed, considering two different contours $\mathcal{C}_1$ and $\mathcal{C}_2$, we have 
\begin{equation}
\mathcal{F}_i(\mathcal{C}_1)-\mathcal{F}_i(\mathcal{C}_2) = 
\iint_{\mathcal{S}(\mathcal{C}_2)\, \backslash\, \mathcal{S}(\mathcal{C}_1)} f_i\,  \text{d}S= 0\, ,
\end{equation}
where the integration is carried out over the surface area $\mathcal{S}(\mathcal{C}_2)\, \backslash\, \mathcal{S}(\mathcal{C}_1)$ delimited by the two contours (see Fig.~\ref{fig:surface}). Therefore,  $\mathcal{F}_{i}(\mathcal{C}_1)=\mathcal{F}_{i}(\mathcal{C}_2)$.

Considering distances sufficiently far from the vortex for the first-order perturbation theory to hold, the momentum-flux tensor can be decomposed as 
\begin{equation}
    \Pi_{ij} =  \bar{\Pi}_{ij} + \delta \Pi_{ij}\, ,
\end{equation}
where $ \delta \Pi_{ij}$ denotes a small disturbance of the uniform background momentum-flux tensor $\bar{\Pi}_{ij} $. Similarly, any quantity $y$ will be expanded to first order as $y = \bar{y} + \delta y$, where $\delta y$ denotes a small disturbance of the uniform background quantity $\bar{y}$. Since the force for the unperturbed uniform background flows must evidently vanish by symmetry, the corresponding force in the presence of the vortex~\eqref{jouko-def} will be given to first order by \begin{equation}
\label{jouko-def2}
\mathcal{F}_{i} = \oint_\mathcal{C}  \delta \Pi^j_{i} \, \alpha_j \, \text{d}\ell\, , 
\end{equation}
see also \cite{carter2002relativistic, carter2005covariant}.

\subsection{Momentum-flux tensor for npe-matter}

Let us assume that the vortex and the $N_{\p}$ fluxoids pinned to it are evolving in a cold\footnote{The temperature in mature neutron stars is expected to be low enough for thermal excitations to be negligible, see, e.g.,~\cite{potekhin2015} for a review and~\cite{beloin2018} for recent neutron-star cooling simulations.} mixture of superfluid neutrons, superconducting protons and degenerate electrons. Such conditions are expected to be met in the outer core of neutron stars. The basic model of such a three-component superconducting-superfluid mixture we adopt here has been described by \cite{carter1998}. Although developed in the relativistic context, their covariant approach remains formally applicable to the Newtonian spacetime since it is based on Cartan's exterior calculus (see also \cite{carter2004,carter2005covariant,carterchamel2005} for the fully 4D covariant nonrelativistic formulation and a discussion of the specificity of the Newtonian spacetime). The explicit hydrodynamic equations in the usual 3+1 spacetime decomposition and based on a similar convective variational action principle have been derived by \cite{prix2004variational,prix2005var}. 

The momentum-flux tensor can be decomposed as 
\begin{equation}
    \Pi_{ij} = \Pi_{ij}^\text{(nuc)} +  \Pi_{ij}^\text{(e)} +  \Pi_{ij}^{\text{(em)}} \, ,
\end{equation}
where $\Pi_{ij}^\text{(nuc)}$, $\Pi_{ij}^\text{(e)}$ and $\Pi_{ij}^{\text{(em)}}$ denote the nucleon,  electron and electromagnetic contributions, respectively. 
The nucleon part reads~\citep{carter1998}
\begin{equation}
\label{eq:nuc_stress_tensor}
   \Pi_{i}^{\text{(nuc)}j}= n_{\n}^j\,  \pi^{\n}_i + n_{\p}^j\, \left(\pi^{\p}_i - \frac{e}{c} A_i\right) +  \Psi \, \delta^j_{i}\, , 
    \end{equation}   
where $n_{\n}^i =  n_{\n} v_{\n}^i$ and $n_{\p}^i = n_{\p} v_{\p}^i$ denote the neutron and proton currents respectively, $\pi^{\n}_i$ and $\pi^{\p}_i$ stand for the associated generalised momenta per particle, $A_i$ is the magnetic potential vector,   
$\Psi$ is the generalised pressure of the nucleon mixture, and $\delta_{i}^{j}$ is the Kronecker symbol. The generalised momenta $\pi^{\n}_i$ and $\pi^{\p}_i$ are related to the purely nuclear momenta  $p^{\n}_i$ and $p^{\p}_i$ (as obtained in the absence of electromagnetic fields) by the following relations~\citep{carter1998}
\begin{equation}
    \pi^{\n}_i = p^{\n}_i\, , \ \ \pi^{\p}_i = p^{\p}_i +  \dfrac{e}{c}A_i\,  .
\end{equation}
As stressed by \cite{carter1989}, the distinction between momenta and currents is crucial. Because neutrons and protons are strongly interacting, they are mutually coupled by nondissipative entrainment effects of the kind originally discussed in the context of superfluid $^3$He$-^4$He mixtures by \cite{andreev1976three} such that the nucleon momenta are expressible as~\citep{carter1998}
\begin{align}
\label{eq:entrainment}
 p_i^{\n} = &\  \gamma_{ij}\left(\mathcal{K}^{\, \nn}\,  n_{\n}^j +\mathcal{K}^{\, \np} \, n_{\p}^j \right)\, , \nonumber \\
 p_i^{\p} = &\ \gamma_{ij}\left(\mathcal{K}^{\, \pn}\,  n_{\n}^j +\mathcal{K}^{\, \pp}\,  n_{\p}^j \right)\, ,
\end{align}
where $\gamma_{ij}$ denotes the space metric. The entrainment coefficients $\mathcal{K}^{\, \nn}$, $\mathcal{K}^{\, \np}=\mathcal{K}^{\, \pn}$, and $\mathcal{K}^{\,  \pp}$ are not all independent since Galilean invariance imposes the following relations:
\begin{align}
\label{eq:kappa}
\mathcal{K}^{\, \nn} \,  n_{\n} +\mathcal{K}^{\, \np}\,  n_{\p} & = m_{\n}\, , \nonumber \\
\mathcal{K}^{\, \pn}\,  n_{\n} +\mathcal{K}^{\, \pp} \, n_{\p}&  = m_{\p}\, .
\end{align}
The entrainment coefficients depend on the baryon number density and on the composition (see, e.g., \cite{gusakov2005,chamel2008two,kheto2014,sourie2016numerical}). They may also depend on the relative nucleon currents so that the relations~\eqref{eq:entrainment} between nucleon momenta and currents are not necessarily linear~\citep{leinson2017,leinson2018}.

The electromagnetic momentum-flux tensor is given by the usual expression (in Gaussian cgs units)
\begin{equation}
\label{eq:Pi_EM}
\Pi_{i}^{\text{(em)}j}  = -\dfrac{1}{4\pi}\left( E_iE^j + B_iB^j- \dfrac{1}{2}E^k  E_k\delta_{i}^{j} -\dfrac{1}{2}B^k B_k\delta_{i}^{j} \right)\, , \end{equation}
where $E^i$ and $B^i$ denote the electric and magnetic fields.

At the vortex scale, electrons do not form a fluid\footnote{We recall here that we are dealing with scales large compared to the size of the vortex-fluxoid configuration, but small with respect to the typical intervortex distance $d_{\n} \simeq 1/ \sqrt{\mathcal{N}_{\n}}\sim 10^{-3}$~cm. The electron mean free path is generally much larger than $d_{\n}$, see, e.g., \cite{shternin2008shear,glampedakis2011magnetohydro}.} but are in a ballistic regime following classical trajectories. Instead of following the purely hydrodynamic treatment of \cite{carter1998} for the electron momentum-flux tensor $\Pi_{ij}^\text{(e)}$, we adopt here the expression given by Eq.~(23) of~\cite{gusakov2019}. 

The total force~\eqref{jouko-def2} experienced by the vortex can thus be decomposed as  
\begin{equation}
\label{eq:force_decomp}
\mathcal{F}_{ i} = \mathcal{F}^{\text{(nuc)}}_{ i} + \mathcal{F}^\text{(e)}_{ i}+ \mathcal{F}^\text{(em)}_{ i}\, , 
\end{equation}
where the separate contributions 
\begin{equation}
\label{eq:force_nuc}
    \mathcal{F}^{\text{(nuc)}}_{ i} =    \oint_\mathcal{C} \delta \Pi^{\text{(nuc)} j}_{i}  \, \alpha_j \, \text{d}\ell\, , 
\end{equation}
\begin{equation}
\label{eq:force_e}
    \mathcal{F}^{\text{(e)}}_{ i} =    \oint_\mathcal{C} \delta \Pi^{\text{(e)} j}_{i}  \, \alpha_j \, \text{d}\ell\, , 
\end{equation}
and 
\begin{equation}
\label{eq:force_em}
    \mathcal{F}^{\text{(em)}}_{ i} =    \oint_\mathcal{C} \delta \Pi^{\text{(em)} j}_{i}  \, \alpha_j \, \text{d}\ell\, , 
\end{equation}
are evaluated in Secs.~\ref{app:nuc_cont},~\ref{app:e_cont} and~\ref{app:EM_cont}, respectively.

\subsection{Nucleon contribution}
\label{app:nuc_cont}

The first-order perturbation in the nucleon momentum-flux  tensor~\eqref{eq:nuc_stress_tensor}  reads
\begin{align}
\label{eq:delta_pi_nuc}
     \delta \Pi^{\text{(nuc)}j}_i = &\ \delta n_{\n}^j \, \bar{\pi}_i^{\n} +
    \bar{n}_{\n}^j \, \delta \pi_i^{\n} + \delta n_{\p}^j \, \bar{\pi}_i^{\p} +
    \bar{n}_{\p}^j \, \left(\delta \pi_i^{\p}-\dfrac{e}{c}\delta A_i\right) \nonumber \\  
    &+ \delta \Psi \, \delta^j_i \, ,
\end{align}
using a gauge such that $\bar{A}_i=0$~\citep{carter2002relativistic}. In the 4D covariant formulations of ~\cite{carter2002relativistic} and~\cite{carter2005covariant}, the first-order perturbation of the nucleon pressure can be expressed as (see Eq.~(A9) of \cite{carterchamel2005})
\begin{equation}
 \label{eq:delta_Pnuc}
    \delta \Psi = - \bar{n}_{\n} \, \delta p_0^{\n} -\bar{n}_{\p} \, \delta p_0^{\p}  - \bar{n}_{\n}^k \, \delta p_k^{\n}  - \bar{n}_{\p}^k \, \delta p_k^{\p} \, , 
\end{equation}
where $p_0^{\n} $ and $p_0^{\p} $ correspond to the time-components of the neutron and proton 4-momenta. These time components can be more explicitly written as 
(see Eq.~(34) of~\cite{prix2005var})
\begin{equation}
\label{eq:p0}
p_0^{_X} = - \mu^{_X} + m_{_X} v_{_X}^2/2 - v_{_X}^k p^{_X}_k\, , 
\end{equation}
with $ \mu^{_X}$ denoting the chemical potential of nucleon species ${\textrm{\scriptsize $X$}}\in\{ \n, \p\}$.
Introducing the time components of the generalised momenta \citep{carter1998}
\begin{equation}
    \pi^{\n}_0 = p^{\n}_0\, , \ \  \pi^{\p}_0 = p^{\p}_0 + e A_0 \, ,
    \label{eq:def_0}
\end{equation}
where $-A_0$ denotes the electric scalar potential, the first-order perturbation in the nucleon momentum-flux  tensor~\eqref{eq:delta_pi_nuc} can be written as
\begin{align}
\label{eq:Pi_nuc_pert}
        \delta \Pi^{\text{(nuc)}j}_i  = &\ \delta n_{\n}^j \, \bar{\pi}_i^{\n} + \bar{n}_{\n}^j \, \delta \pi_i^{\n} + \delta n_{\p}^j \, \bar{\pi}_i^{\p} + 
          \bar{n}_{\p}^j \left(\delta \pi^{\p}_i - \dfrac{e}{c}\delta A_i\right) \nonumber \\
          &- \left(\bar{n}_{\n}  \delta \pi_0^{\n} +\bar{n}_{\p}  \delta \pi_0^{\p}  + \bar{n}_{\n}^k  \delta \pi_k^{\n}  +   \bar{n}_{\p}^k\delta \pi_k^{\p}\right) \delta^j_i  \nonumber \\
          &+ \left( \bar{n}_{\p}  e\, \delta A_0  + \dfrac{e}{c}\bar{n}_{\p}^k \delta A_k\right) \delta^j_i \, .
\end{align}
The nucleon force~\eqref{eq:force_nuc} acting on the vortex can thus be decomposed as 
\begin{equation}
\label{eq:Fnuc}
\mathcal{F}^\text{(nuc)}_{ i} = \mathcal{F}_{\text{E}\ i} + \mathcal{F}_{\text{t}\ i}+ \mathcal{F}_{\text{Mn}\ i} + \mathcal{F}_{\text{Mp}\ i} +  \mathcal{F}_{\Phi\ i} \, , 
\end{equation}
where the different force terms are given by
\begin{equation}
\label{eq:FA}
\mathcal{F}_{\text{E}\ i} = - \oint_\mathcal{C} \left(
\bar{n}_{\n} \, \delta \pi_0^{\n} +\bar{n}_{\p} \, \delta \pi_0^{\p} - \bar{n}_{\p} \, e\, \delta A_0\right)\alpha_i  \, \text{d}\ell\, , 
\end{equation}
\begin{equation}
\label{eq:FB}
\mathcal{F}_{\text{t}\ i} = \oint_\mathcal{C} \left(  \delta   n^j_{\n} \, \bar{\pi}^{\n}_i  + \delta n^j_{\p} \, \bar{\pi}^{\p}_i \right)\, \alpha_j \, \text{d}\ell\, , \end{equation}
\begin{equation}
\label{eq:FC}
\mathcal{F}_{\text{Mn}\ i} = \oint_\mathcal{C} \left(   \bar{n}_{\n}^j\, \delta \pi^{\n}_i  - \delta^j_i \bar{n}_{\n}^k \, \delta \pi^{\n}_k\right)\alpha_j  \, \text{d}\ell\, , 
\end{equation}
\begin{equation}
\label{eq:FD}
\mathcal{F}_{\text{Mp}\ i} = \oint_\mathcal{C}  \left(   \bar{n}_{\p}^j\, \delta \pi^{\p}_i  - \delta^j_i \bar{n}_{\p}^k \, \delta \pi^{\p}_k\right)\alpha_j  \, \text{d}\ell\, , 
\end{equation}
and 
\begin{equation}
\label{eq:FE}
\mathcal{F}_{\Phi\ i} = -\dfrac{e}{c}\oint_\mathcal{C}  \left(   \bar{n}_{\p}^j\, \delta A_i  - \delta^j_i \bar{n}_{\p}^k \, \delta A_k\right)\alpha_j  \, \text{d}\ell\, .
\end{equation}

Let us first focus on $\mathcal{F}_{\text{E}\ i} $. The equations of motion for the neutron superfluid and for the proton superconductor, as expressed as the vanishing of a suitably generalised vorticity tensor, thus take a very simple form in the 4D covariant approach (see Eqs.~(15), (17) and (18) of \cite{carter1998}, or Eqs.~(161) and (171) of \cite{carter2004}). In  the usual spacetime decomposition, the stationary limit of these equations reduces to $\nabla_i \, \pi^{\n}_0 = \nabla_i \, \pi^{\p}_0  = 0$. As shown in Appendix~\ref{app:eom}, this result can also be obtained from Eqs.~(26)$-$(29) of~\cite{prix2005var} or from Eqs.~(B4) and (B5) of~\cite{gusakov2019} in the particular case in which mutual neutron-proton entrainment effects are neglected. The above equations  imply that both $\pi^{\n}_0$ and $\pi_0^{\p}$ are uniform, i.e., $\pi^{\n}_0 = \bar{\pi}^{\n}_0$ and $\pi^{\p}_0 = \bar{\pi}^{\p}_0$, or equivalently, $\delta \pi^{\p}_0= \delta \pi^{\n}_0 =0$. Therefore, we get
\begin{equation}
    \label{eq:FA_2}
    \mathcal{F}_{\text{E}\ i} =  \bar{n}_{\p} \, e \oint_\mathcal{C} 
 A_0\, \alpha_i  \, \text{d}\ell\, ,
\end{equation}
using the fact that $\delta A_0 = A_0-\bar{A}_0 = A_0$ since the uniform background value vanishes (in an appropriate gauge). Equation~\eqref{eq:FA_2} can thus be interpreted as the (opposite of the) force acting on charged protons due to the electric field. 

Introducing the coefficients $\mathcal{D}_{\n}$ and  $\mathcal{D}_{\p}$ as
\begin{equation}
   \mathcal{D}_{\n} = \oint_\mathcal{C} n_{\n}^j \, \alpha_j  \, \text{d}\ell \ \ \text{and} \ \  \mathcal{D}_{\p} = \oint_\mathcal{C} n_{\p}^j \, \alpha_j  \, \text{d}\ell\, ,
\end{equation}
and recalling that $\bar{p}^{_X}_i = \bar{\pi}^{_X}_i$,  the force term $\mathcal{F}_{\text{t}\ i}$ can be rewritten as 
\begin{equation}
\mathcal{F}_{\text{t}\ i} =  \delta \mathcal{D}_{\n} \, \bar{p}^{\n}_i  +  \delta\mathcal{D}_{\p} \, \bar{p}^{\p}_i \, ,
\end{equation}
where $\delta \mathcal{D}_{\n} = \mathcal{D}_{\n} - \bar{\mathcal{D}}_{\n}$ and  $\delta \mathcal{D}_{\p} = \mathcal{D}_{\p} - \bar{\mathcal{D}}_{\p}$. Using Stokes' theorem, $\mathcal{D}_{\n}$ and $\mathcal{D}_{\p}$ can be equivalently expressed as 
\begin{equation}
\mathcal{D}_{\n}=-\iint_{\mathcal{S}(\mathcal{C})} \nabla_k n_{\n}^k\,    \text{d}S\ \ \text{and} \ \ \mathcal{D}_{\p}=-\iint_{\mathcal{S}(\mathcal{C})} \nabla_k n_{\p}^k \,    \text{d}S\, ,
\end{equation}
where the integrals are over the surface $\mathcal{S}\left(\mathcal{C}\right)$  delimited by the contour $\mathcal{C}$ and d$S$ is the corresponding surface element. Therefore, the background values vanish and  $\delta \mathcal{D}_{\n} = \mathcal{D}_{\n}$ and  $\delta \mathcal{D}_{\p} = \mathcal{D}_{\p}$. The force $\mathcal{F}_{\text{t}\ i}$ is thus associated with transfusive processes, whereby particles of different species are converted into each other by nuclear reactions~\citep{carterchamel2005}. If each species is separately conserved, $\nabla_k n_{\n}^k =0$ and $\nabla_k n_{\p}^k = 0$ must hold everywhere throughout the fluids. In such a case, we deduce that $\mathcal{D}_{\n} = \mathcal{D}_{\p} = 0$, which leads to $\mathcal{F}_{\text{t}\ i} = 0$.

Neutron and proton superflows far from the vortex must obey the irrotationality condition (see Eqs.~(15), (17) and (18) of \cite{carter1998} or Eqs.~(161) and (171) of \cite{carter2004})
\begin{equation}
    \label{eq:irrot}
    \varepsilon^{ijk}\nabla_j \pi^{\n}_k = 0  \ \ \text{and}  \ \  \varepsilon^{ijk}\nabla_j \pi^{\p}_k = 0\, ,
\end{equation}
where $\varepsilon_{ijk}$ is the Levi-Civita symbol. The longitudinal invariance along the vortex, in association with the previous irrotationality conditions, lead to $\hat{z}^j \delta \pi^{\n}_j = 0$ and $\hat{z}^j \delta \pi^{\p}_j = 0$ \citep{carter2005covariant}. Defining $\perp^i_j$ as the operator of projection orthogonal to the vortex, i.e., $\perp^i_j = \delta^i_j - \hat{z}^i\hat{z}_j$, we thus have
\begin{equation}
\perp^j_i \delta \pi^{\n}_j = \delta \pi^{\n}_i\, , \hskip0.5cm \perp^j_i \delta \pi^{\p}_j = \delta \pi^{\p}_i \,  .
\end{equation}
The force term $\mathcal{F}_{\text{Mn}\ i}$~\eqref{eq:FC}  can therefore be recast as 
\begin{equation}
    \mathcal{F}_{\text{Mn}\ i}   = \oint_\mathcal{C} \bar{n}_{\n}^j\delta \pi^{\n}_k \left[ \perp^k_i   \alpha_j -\perp^k_j  \alpha_i \right] \, \text{d}\ell\, .
\end{equation}
Introducing the unit vector $\beta^i$ along the contour such that $\alpha_i = -\varepsilon_{ijk}\beta^j\hat{z}^k$ as illustrated in Fig.~\ref{fig:surface}, and making use of the identity
\begin{equation}
    \perp^k_i\varepsilon_{jlm} - \perp^k_j \varepsilon_{ilm} =\  \perp^k_l \varepsilon_{jim}\, , 
\end{equation}
the force $\mathcal{F}_{\text{Mn}\ i}$ can be equivalently written as
\begin{equation}
    \mathcal{F}_{\text{Mn}\ i}   = - \varepsilon_{ijk} \hat{z}^j \bar{n}_{\n}^k \delta \mathcal{C}^{\n }\, ,
\end{equation}
where the neutron momentum integral $\mathcal{C}^{\n} $ is given by 
\begin{equation}
\mathcal{C}^{\n} =   \oint_\mathcal{C}\pi^{\n}_k \text{d}\ell^k\, ,  
\end{equation}
with $\text{d}\ell^k = \text{d}\ell\, \beta^k$. Since the background value $\bar{\mathcal{C}}^{\n} $ vanishes and using the fact that the neutron momentum circulation integral is quantised, we have
\begin{equation}
  \delta \mathcal{C}^{\n} = \mathcal{C}^{\n} = m_{\n} \, \kappa_{\n}
\end{equation}
 in the presence of a single vortex line. The force term $\mathcal{F}_{\text{Mn}\ i}$ finally reads
\begin{equation}
  \mathcal{F}_{\text{Mn}\ i}   = - \bar{\rho}_{\n}\, \kappa_{\n}\,   \varepsilon_{ijk} \hat{z}^j \bar{v}_{\n}^k \, ,
\end{equation}
where $\rho_{\n} = m_{\n}\,  n_{\n} $. This force can thus be recognized as the Magnus force induced by the (quantised) momentum circulation of the neutron superfluid around the vortex. Similar arguments also apply to the proton superconductor, with the only difference that the proton momentum circulation integral reads
\begin{equation}
   \mathcal{C}^{\p} = \oint_\mathcal{C}\pi^{\p}_k \text{d}\ell^k= m_{\p} \, N_{\p}\, \kappa_{\p}
\end{equation}
with $\kappa_{\p}=h/(2\, m_{\p})$, if $N_{\p}$ fluxoids are enclosed inside the contour. The force term $\mathcal{F}_{\text{Mp}\ i}$ thus reads
\begin{equation}
  \mathcal{F}_{\text{Mp}\ i}   = - \bar{\rho}_{\p}\, N_{\p} \, \kappa_{\p}\,   \varepsilon_{ijk} \hat{z}^j \bar{v}_{\p}^k \, .
\end{equation}

Finally, since the magnetic field $B^i = \varepsilon^{ijk} \nabla_j A_k$ carried by the vortex-fluxoid configuration is directed along $\hat{z}^i$, the $z$-component of the vector potential $A_i$ must vanish, i.e., $A_z = 0$ (see, e.g.,  Eqs.~(7) and (9) of~\cite{gusakov2019}).  As a consequence, we have $\perp^j_i \delta A_j = \delta A_i$. Following a procedure similar to the one used to derive $\mathcal{F}_{\text{Mn}\ i}$ and $\mathcal{F}_{\text{Mp}\ i}$, we obtain 
\begin{equation}\label{eq:FPhi}
  \mathcal{F}_{\Phi\ i}   = \bar{n}_{\p}\dfrac{e}{c} \, \Phi \,   \varepsilon_{ijk} \hat{z}^j \bar{v}_{\p}^k \, ,
\end{equation}
where
\begin{equation}
\label{eq:flux}
   \Phi = \oint_\mathcal{C}A_k \text{d}\ell^k
\end{equation}
is the magnetic flux through the surface $S\left(\mathcal{C}\right)$ delimited by the contour $\mathcal{C}$, and using the fact that the uniform background flux $\bar{\Phi}$ necessarily vanishes. Equation~\eqref{eq:FPhi} can thus be interpreted as the (opposite of the) force acting on charged protons due to the magnetic field.

Collecting terms in Eq.~\eqref{eq:Fnuc}, the nucleon force contribution is finally expressible as
\begin{align}
\label{eq:force_nuc_final}
    \mathcal{F}^\text{(nuc)}_{ i} =  &- \bar{\rho}_{\n}\, \kappa_{\n}\,   \varepsilon_{ijk} \hat{z}^j \bar{v}_{\n}^k  - \bar{\rho}_{\p}\, N_{\p} \, \kappa_{\p}\,   \varepsilon_{ijk} \hat{z}^j \bar{v}_{\p}^k  \nonumber \\
    &+ \bar{n}_{\p} \, e\,  \oint_\mathcal{C} A_0\, \alpha_i  \, \text{d}\ell + \bar{n}_{\p}\dfrac{e}{c} \, \Phi \,   \varepsilon_{ijk} \hat{z}^j \bar{v}_{\p}^k \, .
\end{align}
Let us recall that this expression only holds in the absence of transfusive processes, i.e.,  assuming that each species is separately conserved.

\subsection{Electron contribution}
\label{app:e_cont}

Since electrons are in a ballistic regime at the scale of interest here, the calculation of $\mathcal{F}^{\text{(e)}}_{ i}$ needs a specific treatment, see, e.g., \cite{gusakov2019}. Although~\cite{gusakov2019} mainly focused on a single proton fluxoid, several conclusions of this work are actually very general and can be readily transposed to the vortex-fluxoid configuration under consideration. In particular, the electron force term~\eqref{eq:force_e} can be decomposed into two parts, i.e., 
\begin{equation}
   \mathcal{F}^{\text{(e)}}_{ i}   = \mathcal{F}^{\text{(e,sc)}}_{ i}  + \mathcal{F}^{\text{(e,ind)}}_{ i}  \, ,
\end{equation}
see Eq.~(B27) of~\cite{gusakov2019}, where 
\begin{equation}
    \mathcal{F}^{(\text{e,sc})}_{ i} = \oint_\mathcal{C}\delta \Pi^{(\text{e,sc})j}_i  \alpha_j  \, \text{d}\ell
\end{equation}
is associated with the scattering of electrons off the vortex-fluxoid system and 
\begin{equation}
    \mathcal{F}^{(\text{e,ind})}_{ i} = \oint_\mathcal{C}\delta \Pi^{(\text{e,ind})j}_i  \alpha_j  \, \text{d}\ell
\end{equation}
is an ``induced'' contribution  related to the fact  that the scattered  electrons  carry  a  charge and thus generate a weak electric field far from the vortex. Using (B15) of~\cite{gusakov2019}, this induced contribution reads\footnote{The quantities $e_\text{e}$, $\phi$ and $n_{\text{e}0}$ appearing in Eq.~(B15) of~\cite{gusakov2019} correspond here to $-e$, $-A_0$ and $\bar{n}_\text{e}$, respectively.}
\begin{equation}
\label{eq:F_e_ind}
    \mathcal{F}^{(\text{e,ind})}_{ i} = - \bar{n}_{\e} \, e \oint_\mathcal{C} 
 A_0\, \alpha_i  \, \text{d}\ell\, .
\end{equation}
Besides, the force term $\mathcal{F}^{(\text{e,sc})}_{ i}$ is found to be expressible as
\begin{equation}
    \mathcal{F}^{(\text{e,sc})}_{ i} = D_\text{e} \, \bar{v}_{\text{e}\, i} + D'_\text{e}\,  \varepsilon_{ijk} \hat{z}^j \bar{v}_\text{e}^k \, ,  
\end{equation}
see Eqs.~(29) and (18) of~\cite{gusakov2019}, where $ \bar{v}_{\text{e}}^i$ denotes the asymptotically uniform electron velocity (in the frame where the vortex is at rest). Furthermore,~\cite{gusakov2019} derived the expressions for the coefficients $D_\text{e}$ and $D'_\text{e}$ in the particular context where $\mathcal{F}^{(\text{e,sc})}_{ i}$ is governed by the scattering of electrons off the magnetic field carried by a single fluxoid. His expression for $D'_\text{e}$ (see Eq.~(61) of~\cite{gusakov2019}), i.e.,
\begin{equation}
\label{eq:D_prime_e}
    D'_\text{e} = -\dfrac{e}{c} \bar{n}_\text{e} \Phi \, ,  
\end{equation} 
where $\Phi$ is given by Eq.~\eqref{eq:flux}, happens to be very general in the sense that it does not  depend on the detailed structure of the quantised lines carrying the magnetic flux (as long as electrons follow classical trajectories). Therefore, Eq.~\eqref{eq:D_prime_e} remains valid in the present case where electrons are scattered off the magnetic field carried by the vortex and the pinned fluxoids\footnote{Each proton fluxoid pinned to the vortex carries a quantum of magnetic flux $\phi_0$. Besides, the neutron vortex itself is magnetised due to neutron-proton entrainment effects, and thus carries a  fractional quantum of magnetic flux $\phi_{\n}\simeq - \varepsilon_{\p} \phi_0$, where  $\varepsilon_{\p}$ characterizes the importance of entrainment effects (see, e.g.,~\cite{sedrakian1980mechanism, alpar1984rapid}).}. On the other hand, the expression~(60) of~\cite{gusakov2019} for the drag coefficient $D_\text{e}$ is not applicable here since it depends strongly on the configuration of the quantised line(s) carrying the flux $\Phi$. The determination of  $D_\text{e}$ would thus require (i) to know the exact geometry and structure of the vortex and the $N_{\p}$ fluxoids pinned to it, and (ii) to generalise the microscopic scattering calculations carried out by~\cite{gusakov2019} to the present case.

From the previous considerations, the electron force~\eqref{eq:force_e} finally reads
\begin{equation}
\label{eq:force_e_final}
   \mathcal{F}^{\text{(e)}}_{ i}   = - \bar{n}_{\p} \, e \oint_\mathcal{C} 
 A_0\, \alpha_i  \, \text{d}\ell+  D_\text{e} \,  \bar{v}_{\p i} -\dfrac{e}{c} \bar{n}_{\p} \Phi\,  \varepsilon_{ijk} \hat{z}^j \bar{v}_{\p}^k \, ,  
\end{equation}
where we have made use of the electric charge neutrality condition $\bar{n}_\text{e} = \bar{n}_{\p}$ and the so-called screening condition $\bar{v}_{\text{e}}^i=\bar{v}_{\p}^i$, see,  e.g., \cite{glampedakis2011magnetohydro,gusakov2016}. 

\subsection{Electromagnetic contribution}
\label{app:EM_cont}

As can be seen from Eq.~\eqref{eq:Pi_EM}, the first-order perturbation in the electromagnetic stress tensor $\delta \Pi^{\text{(em)} j}_{i}$ only involves terms of the kind $\delta E^j \bar{E}_i$ and $\delta B^j \bar{B}_i$. However,  the asymptotically uniform magnetic and electric fields must vanish in view of the Meissner effect, i.e., $ \bar{B}_i= 0 $ and $ \bar{E}_i =0$ \citep{carter2002relativistic, gusakov2019}. From  $\delta \Pi^{\text{(em)} j}_{i} =0$, we conclude that
\begin{equation}
    \label{eq:force_EM_final}
    \mathcal{F}^{\text{(em)}}_{ i} = 0 \, .
\end{equation}

\subsection{Final expression and comparison with previous studies}

Combining Eqs.~\eqref{eq:force_nuc_final},~\eqref{eq:force_e_final} and~\eqref{eq:force_EM_final}, the total force per unit length~\eqref{eq:force_decomp} acting on a neutron vortex to which $N_{\p}$ proton fluxoids are pinned is finally expressible as
\begin{equation}\label{eq:force-final}
    \mathcal{F}_i = - \bar{\rho}_{\n}\, \kappa_{\n}\,   \varepsilon_{ijk} \hat{z}^j \bar{v}_{\n}^k  - \bar{\rho}_{\p}\, N_{\p} \, \kappa_{\p}\,   \varepsilon_{ijk} \hat{z}^j \bar{v}_{\p}^k +  D_\text{e} \,  \bar{v}_{\p i}\, .
\end{equation}
Note that the electron force proportional to $D_\text{e}^\prime$ derived by \cite{gusakov2019}, i.e., the last term in Eq.~\eqref{eq:force_e_final}, is exactly cancelled by the proton force $\mathcal{F}_{\Phi \ i}$. Let us remark that Eq.~\eqref{eq:force-final} is also applicable to determine the force acting on a neutron-proton vortex cluster of the kind proposed by \cite{sedrakian1995superfluid}.

In the absence of pinning ($N_{\p} = 0$), a situation considered by \cite{alpar1984rapid}, a neutron vortex still experiences a drag force due to the scattering of electrons off the magnetic field induced by the circulation of entrained protons. By setting $N_{\p} = 1$ and $\kappa_{\n}=0$, our expression~\eqref{eq:force-final} reduces to that obtained by \cite{gusakov2019} for the force acting on a single fluxoid. For NSs with a rotation period $P_{10}\sim 1$ and a typical magnetic field $B_{12}\sim 1$, the number of pinned fluxoids may potentially be as large as $N_{\p}\sim \mathcal{N}_{\p}/\mathcal{N}_{\n}\sim 10^{13}$ so that the second term in Eq.~\eqref{eq:force-final} may have a very strong impact on the vortex dynamics. To illustrate the relative importance of the different force terms on the vortex motion, we give the expression of the vortex velocity in the next section. 

\subsection{Vortex motion} 

Once expressed in a frame where the neutron vortex moves at the velocity $v_{\mathrm{L}}^i$, the vortex motion can be obtained from the force balance equation $\mathcal{F}_i  = 0$ (neglecting the masses of the different quantised lines as in previous studies). Following the classical approach of \cite{hall1956vinen} and  considering velocities orthogonal to $\hat{z}^i$, the vortex velocity is found to be given by 
 \begin{align}
 \label{eq:vortex-vel}
v_{\mathrm{L}}^i &= \bar{v}_{\p}^i + \mathcal{B}\, \varepsilon^{ijk}\hat{z}_j \bar{w}_{\p\!\n\, k} + \left(1-\mathcal{B}'\right)\varepsilon^{ijk}\hat{z}_j\varepsilon_{klm}\hat{z}^l \bar{w}_{\p\!\n}^m  \nonumber \\ 
&= \bar{v}_{\n}^i + \mathcal{B}\, \varepsilon^{ijk}\hat{z}_j \bar{w}_{\p\!\n\, k} - \mathcal{B}'  \varepsilon^{ijk}\hat{z}_j\varepsilon_{klm}\hat{z}^l \bar{w}_{\p\!\n}^m\, , 
\end{align}
where $ \bar{w}_{\p\!\n}^i = \bar{v}_{\p}^i - \bar{v}_{\n}^i $ denotes the relative velocity far from the quantised lines, see Appendix~\ref{app:vortex_vel} for details (the coefficients usually denoted by $ \mathcal{B}$ and $ \mathcal{B}'$ in the neutron-star literature were indicated by $\alpha$ and $\alpha'$ in the standard textbook of \cite{donnelly2005}). In this expression, the coefficients $\mathcal{B}$ and $\mathcal{B}'$ are expressible as\footnote{As shown in the accompanying Letter~\citep{sourie20b}, the coefficient $\mathcal{B}$ ($ \mathcal{B}'$) is associated with the dissipative (conservative) contribution in the smooth-averaged mutual-friction force arising at scales large compared to the intervortex spacing.}
\begin{equation}\label{eq:friction-coefs}
    \mathcal{B} = \dfrac{\xi}{\xi^2+\left(1+X\right)^2} \ \  \text{and} \ \    1-\mathcal{B}' = \dfrac{1+X}{\xi^2+\left(1+X\right)^2} \, ,
\end{equation}
where the drag-to-lift ratio $\xi $ and the momentum circulation ratio $X$ read
\begin{equation}
\label{eq:def_xi_X}
    \xi = \dfrac{ D_\text{e}}{\bar{\rho}_{\n} \kappa_{\n}} \ \ \text{and} \ \  X = \dfrac{\bar{n}_{\p}}{\bar{n}_{\n}}  N_{\p} \, .
    \end{equation}
Note that, regardless of the actual values of $\xi$ and $X$, the following inequalities hold
\begin{equation}
\label{conditions}
\mathcal{B} \leq 1/2 \ \ \text{and} \   \  \mathcal{B}' \leq 1\, .
\end{equation}

In the absence of pinning ($N_{\p}=0$), Eq.~\eqref{eq:friction-coefs} reduces to well-known expressions (see, e.g.,~\cite{carter2001}). In particular, the vortex velocity $v_{\mathrm{L}}^i$ coincides with $\bar{v}_{\n}^i$  and $\bar{v}_{\p}^i$ in the weak ($\xi \ll 1$) and strong ($\xi\gg 1$) drag regimes, respectively.  The motion of a vortex is more complicated if proton fluxoids are pinned to it. In particular, the vortex will move with velocity $\bar{v}_{\p}^i$ even in the weak drag limit if the number of pinned fluxoids $N_{\p}$ is large enough such that $X\gg 1$ and $X\gg\xi$. However, it should be stressed that the drag-to-lift $\xi$ itself depends on $N_{\p}$ and may thus be also very large \citep{ding1993magnetic,sedrakian1995superfluid}. Because pinning may lead to a dramatic reduction of the coefficient $\mathcal{B}$, it may also have important implications for the onset of superfluid turbulence, which is thought to be governed by the parameter  $q = \mathcal{B}/\left(1-\mathcal{B}'\right)$~\citep{finne2003}.

\section{Conclusions}

Following an approach originally proposed by Carter and collaborators~\citep{carter2002relativistic,carter2005covariant}, we have derived the  expression for the force per unit length acting on a quantised neutron vortex to which $N_{\p}$ proton fluxoids are attached, see Eq.~\eqref{eq:force-final}. Our expression is very general and can be applied to describe various situations. In particular, Eq.~\eqref{eq:force-final} extends the expression recently obtained by \cite{gusakov2019} for the force per unit length acting on a single fluxoid. 

By clarifying the different contributions to the force, we have shown that the proton-momentum circulation around the vortex induced by the presence of pinned fluxoids gives rise to a Magnus type force. Due to mutual entrainment effects, the distinction between momenta (usually improperly introduced in terms of ``superfluid velocities'') and currents is crucial to obtain the correct expression of the force. Unlike the drag force, this Magnus force does not depend on the microscopic arrangement of pinned fluxoids, as a consequence of the quantisation of the proton circulation.

Although the vortex velocity takes a similar form as in the absence of pinning, see Eq.~\eqref{eq:vortex-vel}, the friction coefficients $\mathcal{B}$ and $\mathcal{B}'$ are found to depend on the dimensionless ratio  $N_{\p}\times \bar{n}_{\p}/\bar{n}_{\n}$ in addition to the drag-to-lift ratio $\xi$. Because $N_{\p}$ may be potentially as large as $\sim 10^{13}$, pinning may have a dramatic impact on the vortex motion and the onset of superfluid turbulence. A major complication comes from the fact that $N_{\p}$, thereby the coefficients $\mathcal{B}$ and $\mathcal{B}'$, may vary along the vortex trajectory: $N_{\p}$ may increase as the vortex moves and encounters more and more fluxoids, but $N_{\p}$ may also decrease as fluxoids get unpinned. The evolution of $N_{\p}$  will generally depend on the spatial distribution of fluxoids in the outer core of a NS, which in turn reflects the geometry of the internal magnetic field. Pinning of neutron vortices to proton fluxoids should thus be taken into account in the modelling of the magneto-rotational evolution of NSs.

\section*{Acknowledgements}

This work was mainly supported by the Fonds de la Recherche Scientifique (Belgium) under grants no. 1.B.410.18F, CDR J.0115.18, and PDR T.004320. Partial support comes also from the COST action CA16214. 

\appendix

\section{Stationary equations of motion for the neutron superfluid and for the proton superconductor}
\label{app:eom}

In this appendix, we show how the stationary equations of motion  for the neutron superfluid and for the proton superconductor, i.e.,
\begin{equation}
\label{eq:eom}
    \nabla_i \, \pi^{\n}_0 = 0 \ \ \text{and} \ \  \nabla_i \, \pi^{\p}_0  = 0\, , 
\end{equation}
can be derived from Eqs.~(26)$-$(29) of~\cite{prix2005var} and from Eqs.~(B4) and (B5) of~\cite{gusakov2019}, although these two studies rely on different approaches.

\subsection{Derivation from Prix (2005)}
\label{app:prix}

Using Eqs.~(26)$-$(29) of~\cite{prix2005var}, the canonical force densities acting on the neutron superfluid and the proton superconductor are respectively given by
\begin{align}
    f^{\n}_i = &\ n_{\n} \left(\partial_t p^{\n}_i - \nabla_i p_0^{\n}\right) - \varepsilon_{ijk} n_{\n} v_{\n}^j \varepsilon^{klm}\nabla_l  p^{\n}_m + n_{\n}m_{\n}\nabla_i\varphi \nonumber\\ &+ \Gamma_{\n} \, \pi^{\n}_i\, , \\
    f^{\p}_i = &\ n_{\p} \left(\partial_t p^{\p}_i - \nabla_i p_0^{\p}\right) - \varepsilon_{ijk} n_{\p} v_{\p}^j \varepsilon^{klm}\nabla_l  p^{\p}_m + n_{\p}m_{\p}\nabla_i\varphi \nonumber\\ &+ \Gamma_{\p} \, \pi^{\p}_i - n_{\p}\, e\, \left(E_i + \dfrac{1}{c}\varepsilon_{ijk} v_{\p}^j B^k\right)\, , 
\end{align}
where $\varphi$ denotes the gravitational gauge field and we have used the shorthand notations
\begin{equation}
\Gamma_{\n} = \partial_t n_{\n} + \nabla_i \left(n_{\n}v_{\n}^i\right)  \ \ \text{and}  \ \ \Gamma_{\p} = \partial_t n_{\p} + \nabla_i \left(n_{\p}v_{\p}^i\right)\, .
\end{equation}
In the absence of external forces, as characterised by $f^{\n}_i =  f^{\p}_i= 0$, and ignoring any transfusive processes, i.e., $\Gamma_{\n}=\Gamma_{\p}=0$, the stationary equations of motion  reduce to 
\begin{align}
    0 = &\ - \nabla_i p_0^{\n}- \varepsilon_{ijk}  v_{\n}^j \varepsilon^{klm}\nabla_l  p^{\n}_m \, , \label{eq:app_prix1} \\
    0 = &\ - \nabla_i p_0^{\p} - \varepsilon_{ijk}  v_{\p}^j \varepsilon^{klm}\nabla_l  p^{\p}_m - e\, \left(E_i + \dfrac{1}{c}\varepsilon_{ijk} v_{\p}^j B^k\right)\, , \label{eq:app_prix2}
\end{align}
where we have neglected the small variations of the gravitational field $\varphi$ on the scales of interest. Making use of the irrotationality conditions~\eqref{eq:irrot}, rewritten as
\begin{equation}
    \varepsilon^{ijk}\nabla_j p^{\n}_k = 0  \ \ \text{and}  \ \  \varepsilon^{ijk}\nabla_j p^{\p}_k + \dfrac{e}{c}B^i= 0\, ,
\end{equation}
in combination with the definition~\eqref{eq:def_0} for $\pi^{\n}_0$ and $\pi^{\p}_0$, Eqs.~\eqref{eq:app_prix1} and~\eqref{eq:app_prix2} reduce to~\eqref{eq:eom}, as expected.

\subsection{Derivation from Gusakov (2019)}

In the alternative approach followed  by~\cite{gusakov2019}, the conservation equations for the neutron and proton momenta read 
\begin{align}
\label{eq:app_gus1}
    \partial_t G_{\n\, i} = &\ - \nabla_k\left(m_{\n} n_{\n} v_{\n}^k v_{\n\, i}\right)-n_{\n}\nabla_i\mu^{\n} \, , \\
    \partial_t G_{\p\, i} = &\ - \nabla_k\left(m_{\p} n_{\p} v_{\p}^k v_{\p\, i}\right)-n_{\p}\nabla_i\mu^{\p} \nonumber\\
    &+ n_{\p} \, e\, \left(E_i + \dfrac{1}{c}\varepsilon_{ijk} v_{\p}^j B^k\right)\, , 
    \label{eq:app_gus2}
\end{align}
where the variations of the gravitational field have been neglected, see Eqs.~(B5) and (B4) of~\cite{gusakov2019}. The quantities $G_{\n\, i} = n_{\n}\,  p^{\n}_i$ and  $G_{\p\, i} = n_{\p} \, p^{\p}_i$ denote the neutron and proton momentum densities, respectively, with $p^{\n}_i= m_{\n} v_{\n\, i}$ and  $p^{\p}_i= m_{\p} v_{\p\, i}$ in the absence of mutual neutron-proton entrainment effects (see Eqs.~\eqref{eq:entrainment} and~\eqref{eq:kappa} with $\mathcal{K}^{\, \np} =0$), as considered in~\cite{gusakov2019}. Ignoring transfusive processes and focusing on stationary situations only, Eqs.~\eqref{eq:app_gus1} and~\eqref{eq:app_gus2} reduce to 
\begin{align}
    0 = &\ - m_{\n} v_{\n}^k \nabla_k v_{\n\, i} - \nabla_i\mu^{\n} \, , \\
    0 = &\ - m_{\p}v_{\p}^k \nabla_k v_{\p\, i} -\nabla_i\mu^{\p} +\, e\, \left(E_i + \dfrac{1}{c}\varepsilon_{ijk} v_{\p}^j B^k\right)\, .
\end{align}
Making use of the relation
\begin{equation}
   A^j\nabla_jA_i  = \dfrac{1}{2}\nabla_i\left(A^jA_j\right) -\varepsilon_{ijk}A^j\varepsilon^{klm}\nabla_lA_m\, , 
\end{equation}
valid for any vector field $A^i$, these equations can then be rewritten as 
\begin{align}
    0 = &\ - \nabla_i \left(  \dfrac{1}{2} v_{\n}^k p^{\n}_k + \mu^{\n}\right) + \varepsilon_{ijk}  v_{\n}^j \varepsilon^{klm}\nabla_l  p^{\n}_m \, , \label{eq:app_gus3}
 \\
    0 = &\ - \nabla_i \left(  \dfrac{1}{2} v_{\p}^k p^{\p}_k + \mu^{\p}\right) + \varepsilon_{ijk}  v_{\p}^j \varepsilon^{klm}\nabla_l  p^{\p}_m \, \nonumber \\&+\, e\, \left(E_i + \dfrac{1}{c}\varepsilon_{ijk} v_{\p}^j B^k\right)\, . \label{eq:app_gus4}
\end{align}
Using Eq.~\eqref{eq:p0}, which can be recast as 
\begin{equation}
   p_0^{_X} = - \mu^{_X} -  m_{_X} v_{_X}^2/2 =- \mu^{_X} - v_{_X}^k p^{_X}_{k}/2\, , 
\end{equation}
in the absence of entrainment effects, Eqs.~\eqref{eq:app_gus3} and~\eqref{eq:app_gus4} are found to be equivalent to (the opposite of) Eqs.~\eqref{eq:app_prix1} and~\eqref{eq:app_prix2}. The irrotationality conditions~\eqref{eq:irrot} therefore lead to Eq.~\eqref{eq:eom}, as in Section~\ref{app:prix}.

\section{Vortex velocity}
\label{app:vortex_vel}

In a frame where the neutron vortex and the $N_{\p}$ fluxoids pinned to it move at the velocity $v_{\mathrm{L}}^i$, the total force per unit length~\eqref{eq:force-final} acting on the quantised lines reads
\begin{align}
    \mathcal{F}_i = &\ - \bar{\rho}_{\n}\, \kappa_{\n}\,   \varepsilon_{ijk} \hat{z}^j \left(\bar{v}_{\n}^k-v_{\mathrm{L}}^k\right)  - \bar{\rho}_{\p}\, N_{\p} \, \kappa_{\p}\,   \varepsilon_{ijk} \hat{z}^j \left(\bar{v}_{\p}^k-v_{\mathrm{L}}^k\right)  \nonumber\\
    &\ +  D_\text{e} \,  \left(\bar{v}_{\p i}- v_{\mathrm{L}\, i}\right)\, .
\end{align}
Neglecting the masses of the quantised lines (as in previous studies), the force balance equation $\mathcal{F}_i  = 0$ leads to 
\begin{equation}
v_{\mathrm{L}}^i = \bar{v}_{\p}^i -\dfrac{1}{\xi}  \varepsilon^{ijk} \hat{z}_j \left(\bar{v}_{\n\, k}-v_{\mathrm{L}\, k}\right) - \dfrac{X}{\xi} \varepsilon^{ijk} \hat{z}_j \left(\bar{v}_{\p\, k}-v_{\mathrm{L}\, k}\right) \, , 
\label{app:vel_1}
\end{equation}
where $\xi$ and $X$ are given by Eq.~\eqref{eq:def_xi_X}. Projecting this latter equation along $\varepsilon_{ijk} \hat{z}^j$ yields 
\begin{align}
    \varepsilon_{ijk} \hat{z}^jv_{\mathrm{L}}^k = &\ \varepsilon_{ijk} \hat{z}^j \bar{v}_{\p}^k -\dfrac{1}{\xi} \varepsilon_{ijk} \hat{z}^j \varepsilon^{klm} \hat{z}_l \left(\bar{v}_{\n\, m}-v_{\mathrm{L}\, m}\right)\nonumber \\ 
    &- \dfrac{X}{\xi} \varepsilon_{ijk} \hat{z}^j \varepsilon^{klm} \hat{z}_l \left(\bar{v}_{\p\, m}-v_{\mathrm{L}\, m}\right) \\
   =  &\ \varepsilon_{ijk} \hat{z}^j \bar{v}_{\p}^k + \dfrac{1}{\xi}  \left(\bar{v}_{\n\, i}-v_{\mathrm{L}\, i}\right) + \dfrac{X}{\xi}  \left(\bar{v}_{\p\, i}-v_{\mathrm{L}\, i}\right) \, , \label{app:vel_2}
\end{align}
where we have used the fact that $\varepsilon_{ijk}\varepsilon^{klm} = \delta^l_i\delta^m_j-\delta^m_i\delta^l_j$, $\delta^i_j$ being the Kronecker delta, and we have only considered velocities orthogonal to $\hat{z}^i$. Using Eq.~\eqref{app:vel_2} in the right-hand side of Eq.~\eqref{app:vel_1} now leads to
 \begin{align}
 v_{\mathrm{L}}^i \left[1+\dfrac{\left(1+X\right)^2}{\xi^2}\right]= &\  \bar{v}_{\p}^i \left[1+\dfrac{X\left(1+X\right)}{\xi^2}\right] +\bar{v}_{\n}^i \dfrac{1+X}{\xi^2}\nonumber\\
 &+ \dfrac{1}{\xi}\varepsilon^{ijk}\hat{z}_j \left(\bar{v}_{\p\, k} -\bar{v}_{\n\, k}\right)\, , 
\end{align}
which can be eventually recast in the equivalent forms~\eqref{eq:vortex-vel}.

\bibliographystyle{mnras}
\bibliography{biblio}

% Don't change these lines
\bsp	% typesetting comment
\label{lastpage}
\end{document}